\documentstyle[12pt]{article}
\begin{document}
\bibliographystyle{unsrt}
\newcommand{\bra}[1]{\left < \halfthin #1 \right |\halfthin}
\newcommand{\ket}[1]{\left | \halfthin #1 \halfthin \right >}
\newcommand{\be}{\begin{equation}}
\newcommand{\ee}{\end{equation}}
\newcommand{\vsig}{\mbox {\boldmath $\sigma$\unboldmath}}
\newcommand{\vep}{\mbox {\boldmath $\epsilon$\unboldmath}}

\title{\bf  The Threshold Pion-Photoproduction Of Nucleons
In The Chiral Quark Model}

\author{Zhenping Li
\\
Physics Department, Carnegie-Mellon University \\
Pittsburgh, PA. 15213-3890 }
\maketitle

\begin{abstract}
In this paper, we show that the low energy theorem (LET)
of the threshold pion-photoproduction can be fully
recovered in the quark model.
An essential result of this investigation is that
the quark-pion operators  are obtained from the effective
chiral Lagrangian, and the low energy theorem does not
require the constraints on the internal structures
of the nucleon. The pseudoscalar quark-pion coupling
generates an additional term at order $\mu=m_{\pi}/M$ only in the
isospin amplitude $A^{(-)}$. The role of the transitions between
the nucleon and the resonance $P_{33}(1232)$ and
P-wave baryons are also discussed, we find that
the leading contributions to the isospin amplitudes
at $O(\mu^2)$ are from the transition between the P-wave
baryons and the nucleon and the charge radius of the nucleon.
The leading contribution from the P-wave baryons only affects
the neutral pion production, and improve the agreement with
data significantly. The transition between the
resonance $P_{33}(1232)$ and the nucleon only
gives an order $\mu^3$ corrections to $A^{(-)}$.
\end{abstract}

\newpage
\subsection*{\bf 1. Introduction}
The pion-photoproduction $N(\gamma,\pi)N$ of the
nucleon is characterized by three independent
amplitudes, $A^{(-)}$, $A^{(+)}$ and $A^{(0)}$.
They are related to the differential cross section for
the threshold pion-photoproduction\cite{dt92} by
\begin{equation}\label{99}
\frac {d\sigma}{d\Omega}=\frac {k_{\pi}}{k} |E_{0+}|^2
\end{equation}
where the amplitudes $E_{0+}$ can be written as
\begin{equation}\label{101}
E_{0+}=\frac{g_{\pi,N}}{8\pi M}
\frac { \left (1+\frac {\mu}2\right )}{(1+\mu)^{\frac 32}}
\chi_f \left ( \frac 12 [\tau_j,\tau_0]A^{(-)}+\tau_j A^{(0)}
+\delta_{j,0}A^{(+)}\right )\chi_i ,
\end{equation}
and $\chi_i$ and $\chi_f$ are initial and final
isospinors respectively. These isospin amplitudes
can be expanded\cite{kr} in terms of the
ratio between the masses of pions and nucleons,
$\mu=\frac {m_{\pi}}{M}$, near the threshold,
and the leading terms are determined by the low
energy theorem(LET)\cite{cgln}, which can be written as
\begin{eqnarray}\label{1}
A^{(-)}=1+O(\mu^2) \nonumber \\
A^{(+)}=A^{(0)}=-\frac {\mu}2 +O(\mu^2).
\end{eqnarray}
This result is model independent, and a direct consequence
of the partial conservation of the axial vector current (PCAC)
and the soft-pion theorem, therefore, it provides an important
test to the models in hadronic physics.  Although the LET only
depends on the static properties of the nucleon and  no
distinction is being made between elementary
and composite particles, the proof of the LET in the quark models
which treat the nucleon as a three quark system is by no means
trivial.  The investigations in both quark models\cite{dt92,dt84,ss87}
 and the relativistic light cone quark model\cite{kc} show that
it is crucial to separate
the center of mass from the internal motions.  However, the
realization of the LET in the nonrelativistic quark model
relies on the constraints on the size and the internal
transitions on the nucleon, although the problem of separating
the center of mass motion does not exist.
  The focus of this paper is to provide a
general proof of the LET in the quark model with no constraint
on the internal structure of the nucleon.
Similar to the Thomson limit and low energy
theorem in the Compton scattering, the LET is determined by the
transition operators corresponding to the center of mass motion,
thus the separation of the center of mass from the internal
motions is essential.  More important, the LET is directly
related to the chiral symmetry, thus the constituent quark
model alone is no longer enough, one should start from the chiral
quark model proposed by Manohar and Georgi\cite{MANOHAR}
which combines the chiral symmetry and the
constituent quarks.

On the other hand, the pion decays of the baryon resonance
play an important role in the $N^*$ physics at CEBAF,
which have been the subject of the many studies\cite{simon}
 in the framework of the constituent quark model, especially
the $^3P_0$ model\cite{LY}.  These calculations
are based mostly on the phenomenological approach,
and few  theoretical justification has been provided,
the LET in the threshold pion photoproduction might be
 a crucial test to the quark-pion couplings in
these phenomenological models,
this is a major motivation of our investigation.
The encouraging result from our study is that the essential
quark-pion coupling operators for  both LET and the
 $^3P_0$ model have exactly the same structure, this
at least shows
that the $^3P_0$ model is consistent with the LET, which has
shown to be quite successful in describing the strong
decays in hadron physics.

In section 3, we shall show that the leading terms from the
internal structure of the nucleon come from the transitions
between the nucleon and the P-wave baryons and the charge radius
of the nucleon at order $\mu^2$, while the
leading term from the transition between the nucleon and
the $\Delta$ resonance, $P_{33}(1232)$, only contributes
to the isospin amplitude $A^{(-)}$ at order $\mu^3$.
This is because the decay of the $\Delta$ resonance
into pion and nucleon is dominated by p-wave which is
suppressed by a factor of $\mu$, and
P-wave baryons decays are dominated by s-wave.
We further find that the leading term in the
transitions between the nucleon and the P-wave baryons
 only contributes to the isospin amplitude $A^{(+)}$,
it will affect the neutral pion production significantly.
This has not been discussed in the literature, while
much discussions have been
concentrated on the nucleon-$\Delta$ transition\cite{ns}.
 Indeed, we find an excellent agreement with data for
$\gamma p\to \pi_0 p$ and $\gamma n\to \pi_0 n$ if the
contribution from the P-wave baryons is included.

The paper is organized as follows; a general derivation
of the LET is given in sections 2, furthermore, we
shall show that the pseudoscalar coupling generates an
additional term  only in $A^{(-)}$ at order $\mu$. The
contributions
from the internal structure of the nucleon is calculated
in the section 3, and the conclusion is given in section 4.

\subsection*{\bf 2. The LET for a three quark system}

In the chiral quark model, the pions are being treated as
Goldstone bosons that generates the spontaneous chiral
symmetry breaking at some scale $\Lambda_{\chi SB}$,
the QCD Lagrangian transforms into an effective chiral Lagrangian
in the low energy limit, which is\cite{MANOHAR};
\begin{equation}\label{2}
{\cal L}=\bar \psi \left [\gamma_{\mu}(i \partial^{\mu} + V^{\mu}
 +A^{\mu}\gamma_5)-m\right ]\psi+\dots
\end{equation}
where the vector and the axial vector currents are
\begin{eqnarray}\label{3}
V_\mu=\frac 12\left (\xi^\dagger \partial_\mu\xi+\xi \partial_\mu
\xi^\dagger\right ) \nonumber \\
A_\mu=\frac 12i\left (\xi^\dagger \partial_\mu\xi-\xi \partial_\mu
\xi^\dagger\right ),
\end{eqnarray}
$\xi=e^{i\pi/f}$, $\pi=\sum_l\phi_l \tau_l$ for the $SU(2)_L
 \otimes SU(2)_R$ and $f_{\pi}\approx 93$ MeV is the pion decay
constant.  The constituent quark mass $m$ in Eq. \ref{2} is
around $350$ MeV due to the chiral symmetry breaking.
Unlike the gluon fields, the chiral transformation of the
electromagnetic gauge field will generate an
additional quark-photon-pion
interaction due to the isospin dependence of the quark charge
\begin{equation}\label{4}
e_q=\frac e2(\frac 13+ \tau_0),
\end{equation}
which is
\begin{equation}\label{5}
H_{\pi,e}=\sum_j \frac {-ie}{2f_{\pi}}
 \left [\tau^j_0, \tau^j_l\right ]
\phi_l\bar \psi_j \gamma_\mu
 \gamma_5 \psi_j A^\mu({\bf k}, {\bf r}_j),
\end{equation}
where $\tau^j_l$ is Pauli isospin matrix, and
$A^\mu({\bf k},{\bf r}_j)$ is a electromagnetic field.
The pion-quark coupling at the tree level is therefore the
standard pseudovector coupling;
\begin{equation}\label{6}
H_{\pi}=\sum_j \frac 1{f_\pi} \bar \psi_j
 \gamma_\mu \gamma_5^j \tau_l^j
\psi_j \partial^{\mu} { \phi_l},
\end{equation}
and the electromagnetic coupling is
\begin{equation}\label{7}
H_{e}=-\sum_j e_j \gamma_\mu^j A^{\mu}({\bf k},{\bf r}_j).
\end{equation}
Therefore, one can write the transition matrix element as
\begin{eqnarray}\label{8}
{\cal M}_{fi}=\langle N_f| H_{\pi,e}|N_i \rangle +
\sum_j\bigg \{ \frac {\langle N_f|H_{\pi} |N_j\rangle
\langle N_j |H_{e}|N_i\rangle }{E_i+\omega-E_j} \nonumber \\
 +\frac {\langle N_f|H_{e}|N_j\rangle \langle N_j|H_{\pi}
|N_i\rangle }{E_i-\omega_{\pi}-E_j}\bigg \},
\end{eqnarray}
where $N_i(N_f)$ is the initial (final) state of the nucleon,
and $\omega (\omega_{\pi})$ represents the
energy of incoming (outgoing) photons(pions).  The first
term in Eq. \ref{8} corresponds to the Seagul
diagram, and the second and the third terms are the direct
and the crossed intermediate resonance pole diagrams.

The nonrelativistic expansion of the quark-photon-pion
interaction gives
\begin{equation}\label{9}
H_{\pi,e}^{nr}=i\sum_j \frac e{2f_\pi}\left
[\tau^j_0,\tau^j_l\right ]\vsig_j \cdot \vep,
\end{equation}
where $\vep$ is the polarization vector of photons,
 and notice that
\begin{equation}\label{10}
\langle N_f| \sum_j \left [\tau^j_0,\tau^j_l\right ]{\vsig_j}
|N_i\rangle =g_A\langle N_f|\left [\tau^T_0,
 \tau^T_l\right ]{\vsig_T}|N_i\rangle ,
\end{equation}
where $\tau^T_l$ and $\vsig^T$ are the total isospin
and spin operators of the nucleon, and $g_A$ is the
axial coupling constant of the nucleon,
we have the expression for the Seagul diagram
\begin{equation}\label{11}
\langle N_f|H_{\pi,e}|N_i\rangle =\frac {ig_Ae}{2f_\pi}
\langle N_f|\left [\tau^T_0,\tau^T_l\right ]{\vsig^T\cdot \vep}
 |N_i\rangle ,
\end{equation}
which only gives the leading term to the isospin
amplitudes $A^{(-)}$.

The calculation of the direct and crossed resonance pole diagrams
follows the similar procedure in Compton Scattering
($\gamma N\to \gamma N$)\cite{BP} in the low energy limit.
Rewritting the electromagnetic interaction $H_{e}$ as
\begin{equation}\label{12}
\langle N_j|H_e |N_i\rangle
 = i(E_j-E_i-\omega) \langle N_j| g_e|N_i\rangle +i\omega \langle N_j
|h_{e}|N_i\rangle
\end{equation}
for the direct pole diagram and
\begin{equation}\label{13}
\langle N_f|H_e |N_i\rangle
 = i(E_f-E_j-\omega) \langle N_f| g_e|N_j\rangle +i\omega \langle N_f
|h_{e}|N_j\rangle
\end{equation}
for the crossed pole diagram, where
\begin{equation}\label{133}
g_e=\sum_j e_j {\bf r}_j \cdot \vep e^{i{\bf k}\cdot {\bf r}_j}
\end{equation}
\begin{equation}\label{14}
h_e=\sum_j e_j {\bf r}_j\cdot \vep (1-\mbox {\boldmath
$\alpha$ \unboldmath}_j
\cdot \hat {\bf k})
e^{i{\bf k}\cdot {\bf r}_j},
\end{equation}
and $\hat {\bf k}=\frac {{\bf k}}{\omega}$,
in which we have replaced the spinor $\bar \psi$ by $\psi^\dagger$ so
that the $\gamma$ matrices are replaced by the matrix $\bf \alpha$,
the second and the third terms in Eq. \ref{8} can be written
as
\begin{eqnarray}\label{15}
{\cal M}_{fi}^{\prime}=i \langle N_f|[g_e,H_{\pi}]|N_i\rangle
+ i\omega \sum_j\bigg \{ \frac {\langle N_f|H_{\pi}
|N_j\rangle \langle N_j |h_{e}|N_i\rangle }{E_i+\omega-E_j}
\nonumber \\  +\frac {\langle N_f|h_{e}|N_j
\rangle \langle N_j|H_{pi}|N_i\rangle }{E_i-
\omega_{\pi}-E_j}\bigg \}.
\end{eqnarray}
The first term in Eq. \ref{15} is also a Seagul term, and
only contributes to the amplitudes $A^{(-)}$, the
nonrelativistic expansion near the threshold gives
\begin{equation}\label{16}
\langle N_f|[g_e,H_{\pi}]|N_i\rangle=-\frac
{\omega_{\pi}g_A}{2f_\pi}\langle N_f| [\tau^T_0,\tau^T_l]
\frac {\vsig^T\cdot {\bf P}_T {\bf R}\cdot\vep+{\bf R}
\cdot \vep \vsig^T \cdot {\bf P}}{2M}|N_i\rangle ,
\end{equation}
in which we have separated the center of mass motion
from the internal motion by
\begin{equation}\label{17}
\frac {{\bf p_j}}{m_j}=\frac {{\bf P}}{M}+\frac
{{\bf p^\prime}_j}{m_j},
\end{equation}
where $P$ is the total momentum of the composite system with mass
$M$ at the position $R$, and ${\bf p}^\prime_j$
represents the internal momentum whose matrix elements
vanish for the initial and final nucleon states.

The nonrelativistic expansion for $h_e$ in Eq. \ref{15}
is\cite{BP,zpli}
\begin{equation}\label{18}
h_e=\sum_j \left [ e_j {\bf r}_j \cdot \vep (1-\frac {{\bf p}_j \cdot
{\bf k}}{m_j \omega})-\frac {e_j}{2m_j}{\vsig_j \cdot (\vep\times \hat
 {\bf k})} \right ],
\end{equation}
which $h_e$ is only expanded to order $1/M$. The corresponding
pion coupling is therefore
\begin{equation}\label{199}
H_{\pi}^{nr} ={-i}\sum_j \frac {\tau^j_l}{f_{\pi}}
\left [ \vsig_j\cdot {\bf k}_{\pi}-\omega_{\pi} \frac {\vsig_j
\cdot {\bf p}_j}{m_j} \right ]
\end{equation}
where $\bf k_{\pi} \approx 0$ and $\omega_{\pi}\approx m_{\pi}$
near the threshold, thus only second term is needed in our
calculation.

In Ref. \cite{zpli93}, we shown how to separate the center of mass
from the internal motion to obtain the Thomson limit and the
low energy theorem in the Compton scattering. Since the incoming
photon energy near the threshold is
\begin{equation}\label{119}
\omega=M\mu \frac {1+\frac {\mu}2}{1+\mu} ,
\end{equation}
the leading terms in the photon energy $\omega$
are also  leading terms in $\mu$.  Replacing the outgoing photon
operator by $H_{\pi}$ in Eq. 22 of Ref. \cite{zpli93}, we have
\begin{eqnarray}\label{19}
{\cal M}_{fi}^\prime  = -\langle N_f|\frac {g_A\omega_{\pi}}{f_{\pi}}
[\tau_0^T,\tau^T_l]\frac { \vsig^T\cdot {\bf P R}\cdot \vep
+{\bf R}\cdot \vep \vsig^T\cdot {\bf P}}{2M}|N_i\rangle \nonumber \\
 + \langle N_f h^c_{\pi}|N \rangle \langle N|\frac {-ie_T{\bf P}
\cdot \vep}{M\omega}+ h^c_{e}|N_i\rangle
- \langle N_f| \frac {-ie_T {\bf P}\cdot \vep}{M\omega}+h^c|N\rangle
\langle N|h^c_{\pi}|N_i\rangle
\end{eqnarray}
where
\begin{equation}\label{20}
h^c_{\pi}=i\sum_j \frac {\omega_{\pi}}{f_\pi} {\tau^j_l}
\frac {\vsig_j\cdot {\bf P}}{M}
\end{equation}
and\cite{zpli93}
\begin{equation}\label{21}
h^c_{e}=e_T {\bf R\cdot \vep}-\mu_N \vsig^T\cdot (\vep
\times \hat {\bf k})
\end{equation}
correspond to the center of mass motion, and the leading terms
for transition operator corresponding to the internal motions
are proportional to $\omega^2$, which will be discussed later.
Therefore, one could use Eq. \ref{10}, and
then use the closure relation since the resulting
transition operator only connects the ground states, we have
\begin{eqnarray}\label{22}
{\cal M}_{fi}^\prime = -\langle N_f|\frac {g_Ae\omega_{\pi}}
{f_{\pi}}[\tau_0^T,\tau^T_l]\frac { \vsig^T\cdot
 {\bf P R}\cdot \vep +{\bf R}\cdot \vep \vsig^T\cdot{\bf P}}
{2M}|N_i\rangle \nonumber \\ -\frac {g_A\omega_{\pi}}{f_{\pi}}
 \langle N_f| \left [\tau^T_l\frac {\vsig^T \cdot {\bf P}}{M},
\quad e_T \frac {-i{\bf P}\cdot \vep}{M\omega}+h^c\right ]
|N_i\rangle .
\end{eqnarray}
The electromagnetic interaction in Eq. \ref{22} is
\begin{equation}\label{23}
e_T\frac {-i{\bf P}\cdot \vep}{M\omega}+h^c=(h^s+\tau_0^T
 h^v)e
\end{equation}
where
\begin{equation}\label{24}
h^{s,v}=\frac 12\frac {-i{\bf P}\cdot \vep}{M\omega}+
\frac 12 {\bf R\cdot \vep} -\mu^{s,v}
\vsig^T\cdot (\vep\times \hat {\bf k}),
\end{equation}
with $\mu^s=\frac 12(\mu_p+\mu_n)$ and $\mu^v=\frac 12(\mu_p-
\mu_n)$.    Using the relation
\begin{equation}\label{25}
[aA, bB]=\frac 12 [a,b]\{ A, B\}+\frac 12\{a,b\} [A,B],
\end{equation}
we obtain the transition amplitudes
\begin{equation}\label{266}
{\cal M}_{fi}=\chi_f \left \{ \frac 12 [\tau^T_l,\tau^T_0]
{\cal M}^{(-)}+\tau^T_l {\cal M}^{(0)}+\delta_{l,0}
{\cal M}^{(+)}\right \}\chi_i ,
\end{equation}
where
\begin{equation}\label{26}
{\cal M}^{(-)}=\frac {eg_A}{f_{\pi}} \langle N_f|\left (-i
\vsig^T\cdot \vep+\mu \left \{ \vsig^T\cdot {\bf P},
\frac 12 {\bf R}\cdot \vep -\frac {-i{\bf P}\cdot
\vep}{2M\omega}-h^v\right \} \right )|N_i\rangle ,
\end{equation}
\begin{equation}\label{27}
{\cal M}^{(+)}=-\frac {e\mu g_A}{f_\pi} \langle N_f|
 \left [\vsig^T\cdot {\bf P}, \frac {-i{\bf P}\cdot
\vep}{2M\omega}+h^v\right ] |N_i \rangle
\end{equation}
and
\begin{equation}\label{28}
{\cal M}^{(0)}=-\frac {\mu eg_A}{f_\pi} \langle N_f|
\left [\vsig^T \cdot {\bf P}, \frac {-i{\bf P\cdot
\vep}}{2M\omega}+h^s\right ] |N_i\rangle .
\end{equation}
In the center of mass frame for the incoming photon and nucleon,
the operator ${\bf P}$ equals to
\begin{equation}\label{29}
{\bf P}=\frac 12 ({\bf P}_i+{\bf P}_f)=-\frac 12 {\bf k},
\end{equation}
it is straightforward to obtain
\begin{equation}\label{30}
A^{(-)}=1+O(\mu^2),
\end{equation}
\begin{equation}\label{31}
A^{(0)}= -\frac 12 \mu+\frac 14 \mu^2
+\frac 14(\kappa_p+\kappa_n) \mu^2
+O(\mu^3)
\end{equation}
and
\begin{equation}\label{32}
A^{(+)}=-\frac 12 \mu
+\frac 14 \mu^2 +\frac 14(\kappa_p-\kappa_n)\mu^2
+O(\mu^3) ,
\end{equation}
where $\kappa$ is the anomalous magnetic moment of the nucleon.
Therefore, we have fully recovered the LET of the threshold
pion-photoproduction.  An interesting result here is that
the isospin amplitude $A^{(-)}$ is only determined by the
Seagul diagram, while $A^{(0)}$ and $A^{(+)}$ are from the direct
and crossed resonance pole diagrams. Unlike the conclusions by the
previous investigations\cite{dt84},  the LET
is determined by the the transition operator corresponding
to the center of mass motion, and no further requirement on
the internal structure of the nucleon is needed. The difference is
that the intermediate states in the low energies are
highly collective, thus the propagator should not be replaced by the
 quark propagator.

The proof of the LET also highlights that it is essential to
start from the effective chiral Lagrangian when pion degrees of
freedom are involved.  The conclusion that the isospin amplitude
$A^{(-)}$ is determined by the Seagul alone is a natural
consequence of the chiral transformation, and can not be
obtained from the direct gauge transformation of the pseudovector
pion-quark coupling operator. Furthermore, there is no free
parameter at the tree level; Eqs. \ref{30}, \ref{31} and
\ref{32} require that the pion nucleon coupling in the
low energy region should be
\begin{equation}\label{33}
g_{\pi N}=\frac {g_A M}{f_{\pi}},
\end{equation}
which is the Goldberger-Treiman relation\cite{gt58}.
Thus, the Goldberger-Treiman relation and the LET
are direct consequences of the chiral symmetry
in the chiral quark model.

Another possible quark-pion interaction is the pseudoscalar
coupling;
\begin{equation}\label{34}
H^{p.s}_{\pi}=\sum_j g_{q\pi}\bar \psi_j \gamma_5^j \psi_j
\tau^j_l\phi_l
\end{equation}
In this case, there is no direct pion-quark-photon coupling
operator, thus, the pion-photoproduction near the threshold
is determined by the direct and crossed resonance pole diagrams.
Taking the same procedure in Eqs. \ref{12} and \ref{13},
 Eq. \ref{15} becomes
\begin{eqnarray}\label{35}
{\cal M}=\langle N_f|[g_e, H^{p.s}_{\pi}]|N_i\rangle +
\sum_j\bigg [\frac {\langle N_f|H^{p.s}_{\pi}|N_j\rangle
\langle N_j|h_e|N_i\rangle }{E_i+\omega-E_j} \nonumber \\ +
\frac {\langle N_f|H^{p.s}_{\pi}|N_j\rangle
\langle N_j|h_e|N_i\rangle }{E_i+\omega-E_j}\bigg ],
\end{eqnarray}
the nonrelativistic expansion of the first term gives
\begin{eqnarray}\label{36}
\langle N_f|[g_e,H^{p.s}_{\pi}]|N_i\rangle
& = & -i\frac {g_{q\pi}}{2}
 \langle N_f|\sum_j [\tau_0^j , \tau^j_l] \frac {
-\vsig_j\cdot {\bf p_j r_j} \cdot \vep+{\bf r_j}\cdot \vep
\vsig_j\cdot {\bf p_j}}{2m_q} |N_i\rangle  \nonumber \\
& =& \frac {g_Ag_{q\pi}}{4m_q} \langle N_f|[\tau^T_0, \tau^T_l]
{\vsig_T\cdot\vep}|N_i\rangle .
\end{eqnarray}
which generates the leading term for the isospin amplitudes
$A^{(-)}$.  Since the nonrelativistic expansion of the pseudoscalar
coupling $H^{p.s}_{\pi}$ has the same expression as the pseudovector
coupling to order $O(1/M)$\cite{LY} except the pion-quark coupling is
replaced by $\frac {g_{q\pi}}{2m_q}$, the second and third terms in
Eq. \ref{35} is the same as those in Eq. \ref{19}, thus, the only
term that will be affected by the pseudoscalar coupling is the
isospin amplitude $A^{(-)}$.  Eq. \ref{26} becomes
\begin{equation}\label{266}
{\cal M}^{(-)}=\frac {eg_{q\pi} g_A}{2m_q} \langle N_f|\left (
\vsig^T\cdot \vep+i\mu \left \{ \vsig^T\cdot {\bf P},
-\frac {-i{\bf P}\cdot \vep}{2M\omega}-h^v\right \}
 \right )|N_i\rangle ,
\end{equation}
while the matrix elements ${\cal M}^{(0)}$ and ${\cal M}^{(+)}$
are the same as those in Eqs. \ref{27} and \ref{28} except
the coupling constant.  The result isospin amplitude $A^{(-)}$
becomes
\begin{equation}\label{377}
A^{(-)}=1+\frac {\mu}2 +O(\mu^2).
\end{equation}
The difference between the pseudovector
and the pseudoscalar couplings is  the term $\frac {\mu}2$ in the
isospin amplitude $A^{(-)}$ in chiral quark model.
 It is well known\cite{BAESNT} that the
Born terms  for the pseudoscalar coupling violates the LET
at order $\mu$, our investigations shows that the pseudoscalar
coupling does generate correct isospin amplitudes $A^{(0)}$
and $A^{(+)}$ if one extends the calculations beyond the
Born approximations.  However, the difference in the isospin
amplitudes in $A^{(-)}$ merits a further study\cite{ss94},
since it has been show\cite{BAESNT} that there should be no
term proportional to $\mu$ in $A^{(-)}$ from the argument
of the crossing symmetry. Thus, the neutral
pion-photoproduction should not be sensitive to the
pseudovector and pseudoscalar couplings.

Even if there is a indeed difference in $A^{(-)}$
for pseudoscalar and pseudovector couplings,
there is no significant phenomenological consequence
if the appropriate parameters are used.
In principle, this difference might be seen in the
ratio between the charged and neutral pion productions,
however, the both charged and neutral pion productions
would affected by the internal structure of the nucleon at
$O(\mu^2)$, which makes it impossible to get a very
accurate and model independent result.

\subsection*{3. The contributions from the intermediate resonances}

The consistent derivation of the LET makes it possible to
discuss the contributions from the internal structure
of the nucleon in this framework.  The most important
contributions from the internal structure of the nucleon
are the charge radius of the nucleon and the transitions
between the nucleon and the P-wave baryons as well as the
$\Delta$ resonance, $P_{33}(1232)$.  The nucleon charge
radius contribution comes from the form factor\cite{zpli93},
$e^{-\frac {\omega^2}{6\alpha^2}}$, for the photon absorptions
near the threshold, expanding this form factor in terms of
$\mu$, we get a correction $-\frac {M^2}{6\alpha^2}\mu^2$
to the isospin amplitude $A^{(-)}$, while the corrections
to  $A^{(0)}$ and $A^{(+)}$ is of order $\mu^3$ since the
leading term of these amplitudes is $-\frac {\mu}2$.

Following the same procedure in discussing the contributions
from the internal transitions to the polarizabilities of
the nucleon, contributions from the transitions between
the nucleon and the excited states can be easily evaluated.
Replacing the electromagnetic
transition operator ${\bf h^*}$ in Eq. 21 of Ref. \cite{zpli93},
the amplitude for the transitions between the nucleon and the
P-wave baryons becomes
\begin{eqnarray}\label{37}
{\cal M}^p=-\frac {3\omega \omega_{\pi}}{f_{\pi}\omega_h}
\langle N_f|  ({\bf h}_{\pi}^3-2{\bf h}_{\pi}^2)\cdot {\bf h_3}
+{\bf h_3}\cdot ({\bf h}_{\pi}^3-2{\bf h}^2_{\pi})|N_i\rangle
\nonumber \\ -\frac {3\omega^2 \omega_{\pi}}{f_{\pi}\omega_h^2}
\langle N_f|  ({\bf h}_{\pi}^3-2{\bf h}_{\pi}^2)\cdot {\bf h_3}
-{\bf h_3}\cdot ({\bf h}_{\pi}^3-2{\bf h}^2_{\pi})|N_i\rangle ,
\end{eqnarray}
where the transition operators ${\bf h_3}$ and
${\bf h}^3_{\pi}-{\bf h}^2_{\pi}$ are
\begin{equation}\label{38}
{\bf h}_3=-\frac 1{\sqrt{3}}\frac {e_3}{\alpha}\vep ,
\end{equation}
\begin{equation}\label{39}
{\bf h}^3_{\pi}-2{\bf h}^3_{\pi}=-i\frac 1{\sqrt{3}}
\frac {\alpha}{m_q}(\tau^3_l \vsig^3-\tau^2_l\vsig^2)
\end{equation}
and the parameter $\omega_h$ is the mass difference between
the nucleon and the P-wave baryons, which is related to the
string constant $\alpha$ by $\omega_h=\frac {\alpha^2}{m_q}$
 in the harmonic oscillator wavefunction.   Therefore, we
can write Eq. \ref{37} as
\begin{equation}\label{40}
{\cal M}^p=-i\frac {\omega\omega_{\pi}}{f_{\pi}\omega_hm_q}
\langle N_f|\left [ \left \{ e^3, \tau^3_l
\right \} { \vsig^3\cdot \vep}-\left \{e^3_0,
\tau^2_l\right \} { \vsig^2\cdot \vep}+\frac
{\omega}{\omega_h}[\tau^3_l,e_3]\vsig^3\cdot
\vep\right ]|N_i\rangle .
\end{equation}
The first two terms in Eq. \ref{40} correspond to the isospin
amplitudes $A^{(0)}$ and $A^{(+)}$ at order $\mu^2$,
while the last term corresponds $A^{(-)}$ and is suppressed
by a factor of $\mu$ relative to the first two terms.
In the $SU(6)$ symmetry limit,
Eq. \ref{40} gives
\begin{equation}\label{41}
{\cal M}^{(0)}=0
\end{equation}
which comes from the symmetry arguments,
\begin{equation}\label{42}
{\cal M}^{(+)}=\frac {4e\mu^2 M^2}{9\omega_h f_{\pi}m_q}
\end{equation}
and
\begin{equation}\label{422}
{\cal M}^{(-)}=\frac {e g_A M^3\mu^3}{3 f_{\pi} \omega_h^2 m_q}.
\end{equation}
Therefore, the leading term from the transition
between the nucleon and the P-wave baryons contributes
to the isospin component $A^{(+)}$, and the contribution
to $A^{(-)}$ is suppressed by a factor of $\mu$.
  Since the ratio $\frac {M^2}{\omega_h m_q}$ is quite large,
this contribution is quite significant, in particular to the $\pi_0$
photo production near the threshold.

For the transition between the nucleon and the resonance
$P_{33}(1232)$, we have
\begin{equation}\label{43}
{\cal M}^{\Delta} =-\frac {\omega\omega_{\pi}}{E_{\Delta}}
\left [ \langle N_f | h^c_{\pi} |\Delta \rangle \langle \Delta
|h^m|N_i \rangle+ \langle N_f |h^m | \Delta \rangle \langle
\Delta |h^c_{\pi} |N_i\rangle \right ],
\end{equation}
where $h^m=\sum_j \frac {e_j}{2m_q}
\vsig_j \cdot (\hat {\bf k} \times \vep)$, and
$E_\Delta$ is the mass difference between the nucleon and
$\Delta$ resonance, and the transition operators
$h^c_{\pi}$ is given in Eq. \ref{20}.
Evaluating Eq. \ref{43} in the $SU(6)$ symmetry limit gives
\begin{equation}\label{44}
{\cal M}^{\Delta}=-\frac {\omega\omega_{\pi}}{E_{\Delta}}
\left [ \langle N_f| \left \{ h^c_{\pi}, h^m\right \}
- \frac {g_A}{f_{\pi}M} \left \{ \mu_N \vsig^T\cdot
(\hat {\bf k}\times \vep), \tau^T_l \vsig^T\cdot {\bf P}\right \}
|N_i\rangle \right ],
\end{equation}
Using the relation for the anticommutator
\begin{equation}\label{45}
\left \{ aA,bB\right \}=\frac 12 \left \{ a,b\right \}\left \{A,B
\right \}+\frac 12 [a,b][A,B]
\end{equation}
and notice that ${\bf P}=-\frac {\bf k}2$ in the center of mass frame,
we have
\begin{eqnarray}\label{46}
{\cal M}^{\Delta}=\frac {\omega \omega_{\pi}}{2f_{\pi}E_{\Delta}M}
\langle N_f| \sum_j [\tau^j_0, \tau^j_l]\frac {i}{2m_q}
\vsig^j \cdot \left ({\bf P}\times (\hat {\bf k}\times \vep)\right )
\nonumber \\
-ig_A(\mu_p-\mu_n) [\tau^T_0, \tau^T_l]
\vsig^T \cdot \left ({\bf P}\times (\hat {\bf k}\times \vep)\right )
|N_i\rangle ,
\end{eqnarray}
 therefore, the transition between the nucleon and the $\Delta$
resonance only contributes to $A^{(-)}$ and it is at order
$O(\mu^3)$, which is
\begin{eqnarray}\label{47}
{\cal M}^{(-)}_\Delta &=& \frac {e g_A\mu^3 M^2}{f_{\pi}E_{\Delta}}
\left [ \frac 1{2m_q}-(\mu_p-\mu_n)\right ]
\nonumber \\ & = & \frac {e g_A\mu^3 M}{2f_{\pi}E_\Delta} \kappa_n .
\end{eqnarray}
It only affects the charged pion production near the threshold.
There is also an additional term from the nucleon-$\Delta$
transition similar to the second term in Eq. \ref{37} by
expanding the propagator in terms of $\mu$, this term
should contribute to the amplitudes $A^{(0)}$ and $A^{(+)}$.
However, it also should be suppressed by a factor of $\mu$
relative to the contribution to $A^{(-)}$, thus should be neglected.
The dependence of the nucleon-$\Delta$ on $\kappa_n$ is
the characteristic  of the electromagnetic
transition between the nucleon and the $\Delta$ resonance,
it is proportional to the magnetic moments of the neutron which
is also the anomalous magnetic magnetic moments of protons and
neutrons in the $SU(6)$ symmetry limit.
The transition between the nucleon and the $\Delta$ resonance
only enters at $O(\mu^3)$ is perhaps
not surprising, the decay of the $P_{33}(1232)$ into
nucleon and pion is in p-wave so that it is of $O(\mu)$
relative to the s-wave. However, since the transition energy
$E_{\Delta}$ is quite small, this contribution should be at
order $O(\mu^2)$ in practical calculations.

One could see the parallel between the contributions of
intermediate resonance states in Compton scattering
$\gamma N\to \gamma N$ and in the threshold
pion-photoproductions;  the transition between the nucleon
and the excited states only enter at $O(\omega^2)$ in the
Compton scattering, which contributes only to the
nucleon polarizabilities, the same is also true for the
threshold pion-photoproduction, and nucleon-$\Delta$
only contributes
to the magnetic polarizabilities which the transition between
the nucleon and the P-wave baryons only contributes to the
electric polarizabilities of the nucleon in the Compton
scattering\cite{zpli93}, while the intermediate $\Delta$
resonance only contributes to the isospin amplitude $A^{(-)}$,
and the transition between the P-wave baryons and the
nucleon contributes isospin amplitudes $A^{(+)}$ only.
The physics is the same, the M1 transition dominates
the nucleon-$\Delta$ transition, and the $E1$ transition
dominates the P-wave baryon nucleon transition.  Since
the threshold pion-photoproduction is dominated by the
E1 transition, it explains why the nucleon-$\Delta$
transition only enter at $O(\mu^3)$, and P-wave baryon
nucleon transition is of order $O(\mu^2)$.  This also
highlights that the P-wave baryon intermediate
states are important for the threshold pion photoproduction,
which has not been discussed fully in the literature.

Since the transition between the nucleon and
the P-wave baryons only contributes to isospin
amplitudes $A^{(+)}$, it is very important in the
neutral pion-photoproduction.  For the process $\gamma n\to \pi_0 n$,
the LET predicts that the amplitude $E_{0+}$ is of order
$\mu^2$, and the transition between the nucleon and
P-wave baryons is also of order $\mu^2$.  This will
modified the results for the neutral pion production
significantly.  Indeed, the experimental results\cite{ps},
although controversial, differ from the LET prediction by a factor
of 2 for neutron targets.  However, if the contributions
 from P-wave intermediate
states are included, the results are very different.
{}From the standard isospin relation
\begin{eqnarray}\label{48}
A(\gamma p\to \pi_0 p)=A^{(+)}+A^{(0)} \nonumber \\
A(\gamma n\to \pi_0 n)=A^{(+)}-A^{(0)},
\end{eqnarray}
we have the total $E_{0+}$ amplitudes for the neutral
pion-photoproduction
\begin{eqnarray}\label{49}
E_{0+}(\gamma p\to \pi_0 p) &=& \frac {e g_A}{8\pi f_{\pi}(1+\mu)}
\left [ -\mu +\frac 12 (1+\kappa_p)\mu^2 +\frac {4 M^2}{9g_A \omega_h
m_q}\mu^2\right ] \nonumber \\
E_{0+}(\gamma n\to \pi_0 n) &=& \frac {e g_A}{8\pi f_{\pi}(1+\mu)}
\left [-\frac 12 \kappa_n \mu^2+\frac {4 M^2}{9g_A \omega_h
m_q}\mu^2\right ]
\end{eqnarray}
For $\omega_h=0.6$ GeV, which is the average mass difference
between the P-wave baryons and the nucleon, $\frac M{m_q}=2.79$
which is the standard value in quark model, $f_{\pi}=93$ MeV,
and $g_A=1.26$, we get
\begin{eqnarray}\label{50}
E_{0+}(\gamma p\to \pi_0 p)=-1.7 \nonumber \\
E_{0+}(\gamma n\to \pi_0 n)=1.0
\end{eqnarray}
in the unit of $10^{-3}/m_{\pi}$, comparing to $-2.4$ for
$\gamma p\to \pi_0 p$ and $0.4$ for $\gamma n\to \pi_0 n$
of the LET predictions.  The contributions from the P-wave
intermediate states increase $E_{0+}$ for $\gamma n\to
\pi_0 n$ by more than factor of 2, and reduce $E_{0+}$
for $\gamma p\to \pi_0 p$ quite significantly.
These improved results are in excellent agreement with
known data, of which the average results are
$-2.0$ for $\gamma p\to \pi_0 p$ and $1.0$ for $\gamma n
\to \pi_0 n$\cite{ps,bc}.

For the isospin amplitude $A^{(-)}$,  the total result is
\begin{equation}\label{55}
{\cal M}^{(-)}=\frac {eg_A}{f_{\pi}} \left [
1-\frac {M^2}{6\alpha^2} \mu^2 +\left (
\frac {M \kappa_n}{2E_{\Delta}}+\frac {M^2}{6\alpha^2}+
\frac {M^3}{3\omega_h^2 m_q}
\right )\mu^3\right ] .
\end{equation}
The contribution from the charge
radius of the nucleon is significant,  if the parameter
$\alpha^2=0.175$ GeV$^2$ is used, it will reduce the charge
pion production by 0.5 $10^{-3}/m_{\pi}$, which is
larger than the previous estimates\cite{ns} of the contribution
from the nucleon-$\Delta$ transitions.  The contributions
from $\Delta$ resonance and the P-wave baryons effectively
cancel each other, so the $\mu^3$ corrections could be neglected.
However, we believe that the order $\mu^3$ corrections is
more model dependent, thus less
reliable because there are many sources that could contribute
to the isospin amplitudes at this order.

\subsection*{4. Conclusion and discussion}

We have shown how  the LET of the threshold pion-photoproduction
is realized in the chiral quark model which the nucleon is treated
as a composite system of three nonrelativistic quarks;  the
LET is only determined by the center of mass motion, and it
requires a quark-pion coupling
from the effective chiral Lagrangian, in particular, the
quark-pion-photon coupling that determines the Seagul term, and
 no constraint on the internal structure of the nucleon is needed.
We also show that the pseudoscalar coupling gives an extra term
$\frac 12 \mu$ in $A^{(-)}$, thus the neutral pion production
is not affected.

The LET provides an important test to the phenomenological
models of the strong decays in hadron physics.
The essential quark-pion operator for the LET is the term that
proportional to $\vsig \cdot {\bf p}$, it is exactly the same
as the operator in the $^3P_0$ model\cite{LY}.  This shows that
$^3P_0$ model is consistent with the LET, thus raises the
possibility to calculate the parameters in the $^3P_0$ model
from the effective chiral Lagrangian.  On the other hand,
the quark-pion coupling operator in
the $^3S_1$ model has different structure,
 thus it could not lead to the LET.  Since $^3S_1$
model could not reproduce the model independent result,
it is theoretically unjustified in this ground.  It is also
interesting to note that studies of the strong decay of $b_1 \to
\omega \pi$ shows that $^3P_0$ model is preferred\cite{LY}.

We show that leading term of the transition between the P-wave
baryons and the nucleon contributes to $A^{(+)}$ at $O(\mu^2)$,
the most crucial test of this term is the neutral pion production,
in particular $\gamma n\to \pi_0 n$, since the leading term
in the LET is also of order $\mu^2$.  Indeed, the result
including the transition between the P-wave baryons and the
nucleon improves the agreement with the known data significantly
for both neutron and proton targets.  We show that the
nucleon-$\Delta$ transition only contributes
to $A^{(-)}$ at order $\mu^3$, and it is less important than the
transition between the nucleon and the P-wave baryons.  This is
because the decay of the $\Delta$ resonance into the nucleon and
pion is  in p-wave, while the P-wave baryons decays are in S-wave,
thus although the $\Delta$ resonance is a very important resonance,
it is suppressed by a factor of $\mu$.  The more important
contribution to the isospin amplitude $A^{(-)}$ is the nucleon
charge radius, which is of $O(\mu^2)$.  We want to emphasize
that contributions from the P-wave baryon intermediate states
and the nucleon charges radius to isospin amplitudes $A^{(-)}$
and $A^{(+)}$ at order $\mu^2$
is model independent, although the numerical
results may vary.  There might be also contributions
from the pion loops, which remains to be studied.
A systematic study of the contributions
at $\mu^3$ may be needed, which the corrections might be
significant in the charge pion productions.
Our results here provide an important challenge to the
experiments; an accurate measurement of the neutral pion
productions will provide a direct probe to the structure of the
nucleon.

Our calculation here also presents a framework to
calculate the $\eta$ and $K$ photoproductions near the
threshold, which the contributions from the resonances
become much more important. The extension of our
study to the pion-electroproduction near the threshold
is also in progress.

The author would like to thank L. Kisslinger who brought my
attention to this problem, and for many useful discussions.
Discussions with B. Keister and S. Scherer are gratefully acknowledged.
This work is supported by U.S. National Science Foundation grant
PHY-9023586.


\begin{thebibliography}{99}
\bibitem{dt92} D. Drechsel and L. Tiator, J. Phys. G: Nucl. Part.
Phys. {\bf 18}, 449(1992).
\bibitem{kr} N. Kroll and M. Ruderman, Phys. Rev. {\bf 93}, (1954).
\bibitem{cgln} G. F. Chew, M. L. Goldberger, F. E. Low and Y. Nambu,
Phys. Rev. {\bf 106}, 1345(1957); S. Fubini, G. Furlan and C. Rossetti,
Nuovo Cimento, {\bf 40}, 1171(1965).
\bibitem{dt84}  D. Drechsel and L. Tiator, Phys. Lett. {\bf 148B},
413(1984).
\bibitem{ss87} S. Scherer, D. Drechsel and L. Tiator, Phys. Lett.
{\bf B193}, 1(1987).
\bibitem{kc} Z-J Cao and L. Kisslinger, Phys. Rev. Lett. {\bf 64},
1007(1990).
\bibitem{MANOHAR} A. Manohar and H. Georgi, Nucl. Phys. {\bf B234},
189(1984).
\bibitem{simon} S. Capstick and W. Roberts, Phys. Rev. {\bf D47},
1994(1993).
\bibitem{LY} Le Yaouanc {\it et al}, Hadron Transitions In The
Quark Model, (Gordon and Breach, New York, 1988).
\bibitem{ns} L. M. Nath and S. K. Singh, Phys. Rev. {\bf C39},
1207(1989); M. G. Olsson and E. T. Osypowski, Nucl. Phys. {\bf B87},
387(1975).
\bibitem{BP} F. E. Close and L. A. Copley, Nucl. Phys. {\bf B19},
477(1970); S. J. Brodsky and J. R. Primack, Ann. Phys. {\bf 52},
315(1969).
\bibitem{zpli}F. E. Close and Zhenping Li, Phys. Rev. {\bf D42},
2194(1990).
\bibitem{zpli93} Zhenping Li, Phys. Rev. {\bf D48}, 3070(1993).
\bibitem{gt58} M. Goldberger and S. Treiman, Phys. Rev. {\bf
110}, 996(1958).
\bibitem{BAESNT} P. De. Baesnt, Nucl. Phys. {\bf B24}, 633(1970).
\bibitem{ss94} S. Scherer, private communications.
\bibitem{ps} X. Pfeil and Y. Schwela, Nucl. Phys. {\bf B45}, 379(1972);
R. L. Crawford and W. T. Morton, Nucl. Phys. {\bf B211}, 1(1983).
\bibitem{bc} F. A. Berends and D. L. Weaver, Nucl. Phys. {\bf B30},
575(1971).
\end{thebibliography}
\end{document}